\documentclass[aps,prl,reprint,superscriptaddress,amsmath,amssymb,showpacs,floatfix,longbibliography]{revtex4-1}

\usepackage{graphicx}
\usepackage{dcolumn}
\usepackage{bm}
\usepackage{color}
\usepackage[utf8]{inputenc}
\usepackage{physics}
\usepackage{mathtools}
\usepackage{soul}
\usepackage{xcolor}

\begin{document}

\title{
Fluctuation-response inequalities for kinetic and entropic perturbations
}

\author{Euijoon Kwon}
\affiliation{Department of Physics and Astronomy \& Center for Theoretical Physics, Seoul National University, Seoul 08826, Republic of Korea}
\affiliation{School of Physics, Korea Institute for Advanced Study, Seoul 02455, Republic of Korea}
\author{Hyun-Myung Chun}
\affiliation{School of Physics, Korea Institute for Advanced Study, Seoul 02455, Republic of Korea}
\author{Hyunggyu Park}
\affiliation{Quantum Universe Center, Korea Institute for Advanced Study, Seoul 02455, Korea}
\author{Jae Sung Lee}
\email{jslee@kias.re.kr}
\affiliation{School of Physics, Korea Institute for Advanced Study, Seoul 02455, Republic of Korea}

\date{\today}

\begin{abstract}
{We derive fluctuation-response inequalities for Markov jump processes that link the fluctuations of general observables to the response to perturbations in the transition rates within a unified framework.
These inequalities are derived using the Cram\'er-Rao bound, enabling broader applicability compared to existing fluctuation-response relations formulated for static responses of current-like observables.
The fluctuation-response inequalities are valid for a wider class of observables and are applicable to finite observation times through dynamic responses.
Furthermore, we extend these inequalities to open quantum systems governed by the Lindblad quantum master equation and find the quantum fluctuation-response inequality, where dynamical activity plays a central role.
}
\end{abstract}

\maketitle

\emph{Introduction.}--The response of a physical system to small perturbations is a fundamental aspect of its behavior~\cite{chaikin1995principles}, and is critical for understanding material properties such as conductivity~\cite{kubo2012statistical} and viscoelasticity~\cite{mason1995optical}.
Near equilibrium, the seminal fluctuation-dissipation theorem states that the system's response is directly related to its spontaneous equilibrium fluctuations~\cite{kubo1966fluctuation}.
Much effort has been devoted to generalizing the connection between response and fluctuations in far-from-equilibrium regimes, expressed as equalities~\cite{agarwal1972fluctuation,speck2006restoring,baiesi2009fluctuations,seifert2010fluctuation,altaner2016fluctuation,owen2020universal,aslyamov2024nonequilibrium}.
More recently, complementary inequalities have been developed to provide upper bounds on  the response in terms of fluctuations~\cite{dechant2020fluctuation,gao2022thermodynamic,chun2023trade,fernandes2023topologically,gao2024thermodynamic,aslyamov2024nonequilibrium,aslyamov2024general,ptaszynski2024dissipation,aslyamov2024nonequilibrium2}.
Although inequalities inherently offer less precise information than equalities, they have practical advantages: they do not require detailed microscopic knowledge of the system, such as the steady-state distribution, which is often necessary for practical application of equalities.

Traditionally, response theory has primarily considered perturbations such as small impulses that alter a given potential~\cite{agarwal1972fluctuation,speck2006restoring,baiesi2009fluctuations}.
In contrast, kinetic perturbations have been largely overlooked, as they affect only reaction rates without changing equilibrium distributions, leading to vanishing responses at equilibirum~\cite{gao2022thermodynamic,chun2023trade,gao2024thermodynamic}.
However, under nonequilibrium conditions, kinetic perturbations become crucial for fully capturing nonequilibrium responses.
Recent studies have established explicit thermodynamic bounds on the kinetic responses of state-dependent observables across various processes~\cite{owen2020universal,gao2022thermodynamic,chun2023trade,fernandes2023topologically,gao2024thermodynamic} and on those of current-like observables in Markov jump processes~\cite{aslyamov2024nonequilibrium,ptaszynski2024dissipation,aslyamov2024nonequilibrium2}.
Among these, the response thermodynamic uncertainty relation (R-TUR)~\cite{ptaszynski2024dissipation} states that the ratio of the kinetic response to the fluctuations of a current-like observable is bounded by the entropy production (EP) rate.
It was found that the R-TUR arises from an identity that connects kinetic response and fluctuations, valid even far from equilibrium~\cite{aslyamov2024nonequilibrium2}.
Although the identities found in~\cite{aslyamov2024nonequilibrium2}, coined fluctuation-response relations (FRRs), can also be used to derive other types of upper bounds on the response to perturbations in the symmetric and anti-symmetric parts of transition rates, their validity is limited to infinitely long observation times.

In this work, we derive inequalities that relate the dynamic response to perturbations in transition rates with the fluctuations of a general observable in Markov jump processes, generalizing the FRRs by encompassing them as a limiting case.
The derivation of these inequalities is based on the Cram\'er-Rao bound, a widely used tool for deriving various uncertainty relations~\cite{dechant2018multidimensional,hasegawa2019uncertainty,lee2019thermodynamic,lee2021universal,Vo2022}.
An important advantage of this method is its simplicity and applicability to finite observation times in steady states, thereby extending the R-TUR for static response~\cite{ptaszynski2024dissipation,aslyamov2024nonequilibrium2} to dynamic response. 
We further extend these inequalities to open quantum systems governed by the Lindblad quantum master equation. 

\emph{Setup.}--We consider a continuous-time Markov jump process governed by the following master equation:
\begin{align} \label{eq:Markov_ori_dyna}
    \dot p_i(t) = \sum_{j (\neq i)} \bm{(} W_{ij} p_j(t) - W_{ji} p_i(t) \bm{)} 
    \;,
\end{align}
where $W_{ij} $ denotes the transition rate from state $j$ to $i$, and $p_i(t)$ represents the probability of the system being in state $i$ at time $t$.
We assume that every transition is bi-directional and that each pair of opposite transitions satisfies local detailed balance for thermodynamic consistency~\cite{seifert2012stochastic,maes2021local}.
The transition rate is parameterized as~\cite{owen2020universal,aslyamov2024nonequilibrium,ptaszynski2024dissipation}
\begin{align}\label{eq:parameterization}
    W_{ij} = \exp \left( B_{ij}(\epsilon) + \frac{F_{ij}(\eta)}{2} \right)
    \;,
\end{align}
where $B_{ij}$ and $F_{ij}$ represent the symmetric and anti-symmetric parts of the transition rate, satisfying $B_{ij} = B_{ji}$ and $F_{ij} = -F_{ji}$, respectively.
Intuitively, in the context of a reaction pathway, $B_{ij}$ represents the energy barrier between two states, while $F_{ij}$ represents the change in entropy due to both the energy difference between the two states and nonequilibrium driving.
We introduce the parameters $\epsilon$ and $\eta$, which control the symmetric and anti-symmetric parts of the transition rates, respectively, without affecting each other.
The steady-state probability of the system, denoted by $\pi_i$, satisfies $\sum_{j(\neq i)} ( W_{ij} \pi_j - W_{ji} \pi_i ) = 0$.
The thermodynamic and kinetic aspects of transitions are characterized by the currents $J_{ij} = W_{ij} \pi_j - W_{ji} \pi_i$ and the traffic $a_{ij} = W_{ij} \pi_j + W_{ji} \pi_i$, respectively, in the steady state.

\begin{figure}
    \includegraphics[width=\columnwidth]{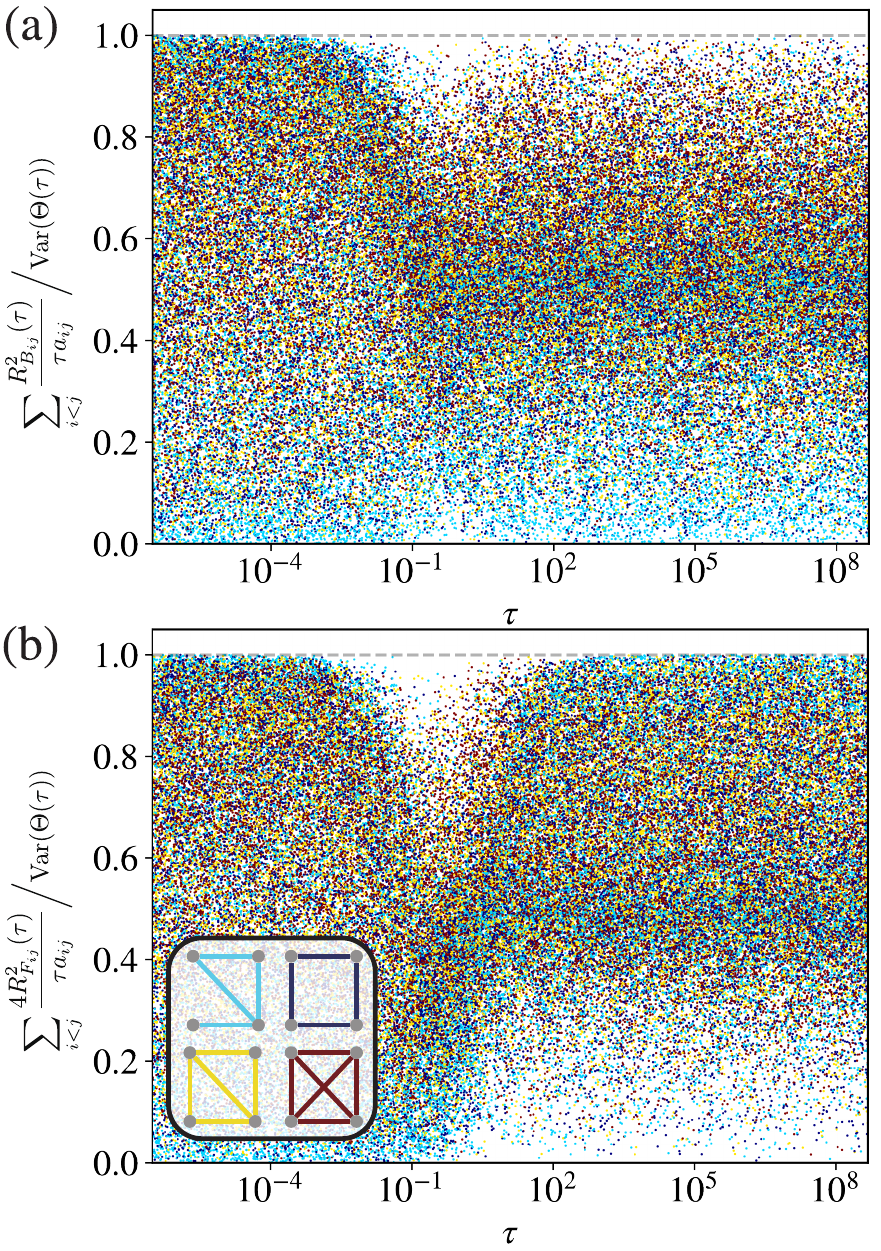}
    \caption{\label{fig:fig_FRI_classical} Numerical verification of FRIs for general observables.
    (a) and (b) correspond to \eqref{eq:FRI_B} and \eqref{eq:FRI_F}, respectively. For the symmetric and anti-symmetric parameters, $B_{ij}$ and $e^{F_{ij}}$ are randomly sampled from $[-2,2]$ and $[0,10]$, respectively. The observation time is given as $\tau = e^x$ where $x$ is drawn randomly from $[-15,20]$. Weights $g_i$ and $\Lambda_{ij}$ are sampled from $[-2,2]$. The network topology is randomly selected from the four possible configurations shown in the inset of (b). Different point colors represent results from the respective topologies, matching the colors in the inset. The total number of points is $10^5$.}
\end{figure}

When the antisymmetric parameter $F_{ij}$ sums to a nonzero value along at least one closed path in the state space, the system is driven out of equilibrium and dissipates energy constantly in the steady state.
This dissipation is characterized by the (mean) EP rate $\dot{\Sigma} = \sum_{i<j} J_{ij} \ln (W_{ij}\pi_j/W_{ji}\pi_i) = \sum_{i<j} J_{ij}F_{ij}$~\cite{seifert2012stochastic}.
We set the Boltzmann constant to unity throughout.
The pseudo-EP rate, a measure of the irreversibility of dynamics, has been found useful in deriving thermodynamic uncertainty relations and is defined as $\dot{\Sigma}_\text{ps} = \sum_{i<j} 2 J_{ij} ^2  / a_{ij}$~\cite{Shiraishi2021,Vo2022,kwon2024unified}. The log-mean inequality, $2/(x+y) \leq (\ln x- \ln y)/(x-y)$ for positive $x$ and $y$, guarantees that $\dot{\Sigma}_\text{ps} \leq \dot{\Sigma}$.
While the EP characterizes the irreversible nature of nonequilibrium systems, the dynamical activity, defined as $\dot{A} = \sum_{i<j} a_{ij}$, serves as a complementary role by describing the time-symmetric aspect of dynamics~\cite{maes2020frenesy}.

To address both the average behavior and fluctuations, we introduce two stochastic quantities: the state identifier $\eta_i(t) = \delta_{s(t),i}$, where $s(t)$ is the state of the system at time $t$, and $N_{ij}(t)$, which denotes the accumulated number of jumps from state $j$ to state $i$ up to time $t$.
To investigate the relations between response and fluctuations, we focus on general time-accumulated observables with arbitrary weights $g_i$ and $\Lambda_{ij}$, defined as
\begin{align}
    \Theta(\tau) = \int_0^\tau dt \left( \sum_{i} g_{i} \eta_{i}(t)
    + \sum_{i\neq j} \Lambda_{ij} \dot{N}_{ij} (t) \right)
    \;,
\end{align}
where $\tau$ is the observation time and $\dot{N}_{ij}(t)$ denotes the rate of change of $N_{ij}(t)$.
We will refer to the observables as current-like if $g_i = 0$ and $\Lambda_{ij} = - \Lambda_{ji}$ for all pairs $(i,j)$, and as state-dependent if $\Lambda_{ij} = 0$ for all pairs $(i,j)$.

\begin{figure}
    \includegraphics[width=\columnwidth]{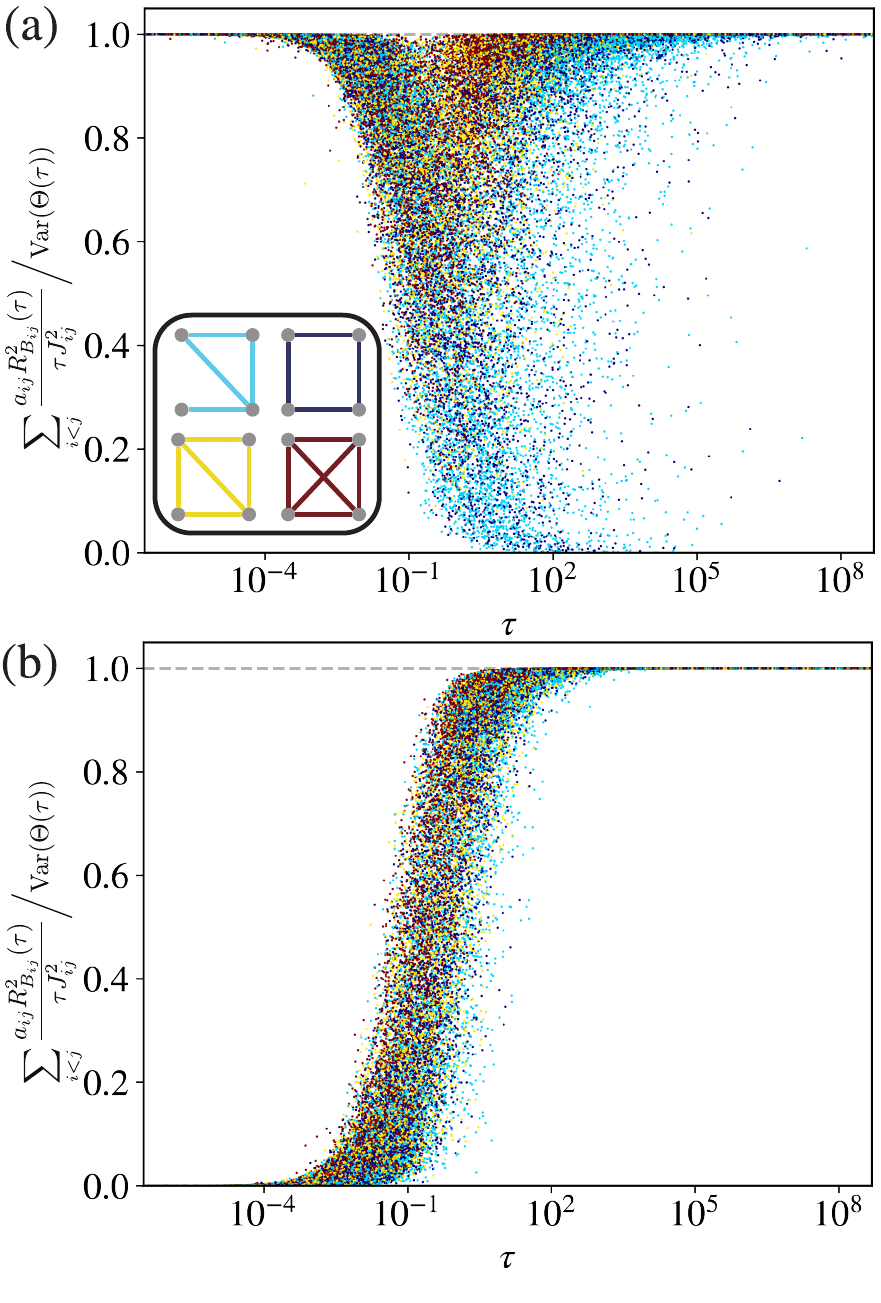}
    \caption{\label{fig:fig_FRI_B2} Numerical verification of FRI~\eqref{eq:FRI_B2} for (a) current-like observables and (b) state-dependent observables. Transition rates, observation times, and weights of observables are sampled within the same ranges as in Fig.~\ref{fig:fig_FRI_classical}. The network topology is randomly selected from the four configurations in the inset of (a), with data point colors matching the respective topologies. The total number of points is $10^5$.}
    \end{figure}

\emph{Fluctuation-Response Inequalities.}--Suppose the system is initially in the steady state for $t<0$, and a system parameter $\theta$ is slightly changed at $t=0$. The transition rates and the probability distributions are then altered as $W_{ij}' = W_{ij} + (\partial_\theta W_{ij})\Delta\theta$ and $p_i(t) = \pi_i + q_i(t) \Delta\theta$, respectively, up to linear order in the change $\Delta \theta$.
The mean value of the observable $\Theta(\tau)$, measured from $t=0$, deviates from the unperturbed value by $\Delta \Theta(\tau) = \langle \Theta(\tau) \rangle - \langle \Theta(\tau) \rangle_0$, where $\langle \bullet \rangle$ and $\langle \bullet \rangle_0$ denote the ensemble averages over perturbed and unperturbed dynamics, respectively.
We define the dynamic response with respect to the change in $\theta$ as $R_\theta(\tau) = \lim_{\Delta \theta\to 0} \Delta\Theta(\tau)/\Delta \theta$.
The Cram\'{e}r-Rao bound provides a general relation between the response to the perturbation and the variance of the observable as $R_\theta^2(\tau) \leq {\rm Var}{\bm({}} \Theta(\tau) {\bm{)}} \mathcal{I}_\theta(\tau)$, where $\mathcal{I}_\theta(\tau) = -\langle \partial_\theta^2 \ln \mathcal{P}[\{s(t)\}_{t=0}^\tau] \rangle_0$ is the Fisher information of the path probability $\mathcal{P}[\{s(t)\}_{t=0}^\tau]$ for the unperturbed dynamics~\cite{hasegawa2019uncertainty,dechant2018multidimensional}, and ${\rm Var} ( \bullet )$ denotes the variance calculated in unperturbed dynamics.
It is important to note that the derivative $\partial_\theta$, used only for notational brevity in the Fisher information, does not apply to the initial distribution since the perturbation does not alter the initial condition.

For multiple perturbation parameters $( \theta_1, \cdots, \theta_K )$, the Cram\'{e}r-Rao bound generalizes to $\sum_{\alpha,\beta} R_{\theta_\alpha}(\tau) [\mathcal{I}^{-1}(\tau)]_{\theta_\alpha \theta_\beta} R_{\theta_\beta}(\tau) \leq {\rm Var} {\bm(} \Theta(\tau) {\bm)}$ with the Fisher information matrix whose elements are given by $\mathcal{I}_{\theta_\alpha \theta_\beta}(\tau) = -\langle \partial_{\theta_\alpha} \partial_{\theta_\beta} \ln \mathcal{P}[\{s(t)\}_{t=0}^\tau] \rangle_0$~\cite{kay1993fundamentals}.
When the set of perturbation parameters consists of
either the symmetric parameters $ B_{ij}$ or the anti-symmetric parameters $F_{ij}$, the Fisher information matrix becomes diagonal, leading to the following inequalities (see Appendix):\begin{align}\label{eq:FRI_B}
    \sum_{i<j} \frac{R_{B_{ij}}^2(\tau)}{\tau a_{ij}} & \leq {\rm Var} {\bm(} \Theta(\tau) {\bm )} \;,
\end{align}
\begin{align}\label{eq:FRI_F}
    \sum_{i<j} \frac{4R_{F_{ij}}^2(\tau)}{\tau a_{ij}} & \leq {\rm Var} {\bm (} \Theta(\tau) {\bm )} \;,
\end{align}
where the summation is taken only over pairs $i<j$ due to the conditions $B_{ij}=B_{ji}$ and $F_{ij}=-F_{ji}$.
We will refer to perturbations in $B_{ij}$ and $F_{ij}$ as kinetic and entropic perturbations, respectively. 
This terminology is motivated by the fact that, in relaxation dynamics to equilibrium, the former changes the time scale of relaxation without affecting the equilibrium distribution, whereas the latter modifies the entropy of the system at the equilibrium.
For observables in the set $\mathcal{S}$, which includes current-like observables, state-dependent observables, and their combinations, the dynamic response to the symmetric parameter $B_{ij}$ and that to the anti-symmetric parameter $F_{ij}$ are interrelated by the identity $R_{B_{ij}}(\tau)/R_{F_{ij}}(\tau) = 2J_{ij}/a_{ij}$ for all observation times $\tau$ (see Appendix).
Plugging this identity into \eqref{eq:FRI_F}, we obtain another inequality involving the kinetic response:
\begin{align}\label{eq:FRI_B2}
    \sum_{i<j} \frac{a_{ij}R_{B_{ij}}^2(\tau)}{\tau J_{ij}^2} & \leq {\rm Var} {\bm(} \Theta(\tau) {\bm )} ~~~ {\rm for} ~ \Theta \in \mathcal{S} \;.
\end{align}
We will refer to the inequalities (\ref{eq:FRI_B}-\ref{eq:FRI_B2}) as fluctuation-response inequalities (FRIs) following~\cite{dechant2020fluctuation}, where a general inequality between fluctuations and response is proposed based on an information-theoretic approach.

The FRIs generalize the FRRs discovered in~\cite{aslyamov2024nonequilibrium2} in two ways.
First, these inequalities are valid for all observation times $\tau$ and encompass the FRRs as the dynamic response function $R_\theta(\tau)$ converges to the static response function in the limit $\tau\to \infty$.
Second, unlike the FRRs, which is applicable only to current-like observables, the FRIs allow general observables for \eqref{eq:FRI_B} and \eqref{eq:FRI_F}, and observables in the set $\mathcal{S}$ for \eqref{eq:FRI_B2}.
While the FRRs are equalities derived from extensive linear algebraic steps, the FRIs are inequalities resulting from a straightforward application of the Cram\'{e}r-Rao bound.

Figure~\ref{fig:fig_FRI_classical} illustrates the validity of the FRIs \eqref{eq:FRI_B} and \eqref{eq:FRI_F} in four-state Markov jump processes with various topologies and system parameters. The vertical axes in all figures represents $\mathcal{Q}$, a quantity obtained by transforming inequalities into the form $\mathcal{Q} \leq 1$.
Unlike current-like observables as reported in~\cite{aslyamov2024nonequilibrium2}, the FRIs \eqref{eq:FRI_B} and \eqref{eq:FRI_F} do not appear to converge to equalities in the limit $\tau \to \infty$ for general observables.
The validity of FRI \eqref{eq:FRI_B2} for current-like and state-dependent observables is examined in Fig.~\ref{fig:fig_FRI_B2}.
The numerical results suggest that the FRI \eqref{eq:FRI_B2} converges to equality in both limits $\tau \to 0$ and $\tau \to \infty$ for current-like observables, and only in the limit $\tau \to \infty$ for state-dependent observables.
While the convergence to equality in the limit $\tau \to 0$ is straightforward to verify (see Appendix), verifying the convergence in the limit $\tau \to \infty$ requires a more sophisticated analysis, as performed in~\cite{aslyamov2024nonequilibrium2}.

\emph{Response Uncertainty Relations.}--Further applications of the Cauchy-Schwartz inequality to the FRIs lead to the recently discovered R-TUR and its variants~\cite{ptaszynski2024dissipation,aslyamov2024nonequilibrium2}.
The responses of interest are now $R_\epsilon(\tau) = \sum_{i<j} b_{ij} R_{B_{ij}}$ and $R_\eta(\tau) = \sum_{i<j} f_{ij} R_{F_{ij}}$, with the shorthand notations $b_{ij} = d_\epsilon B_{ij}$ and $f_{ij} = d_\eta F_{ij}$.
Applying the Cauchy-Schwartz inequality $\sum_{i<j} (x_{ij}/y_{ij})^2 \geq ( \sum_{i<j} x_{ij} )^2/(\sum_{i<j} y_{ij}^2)$ to \eqref{eq:FRI_B} and \eqref{eq:FRI_F}, we obtain the following results: choosing $x_{ij} = b_{ij} R_{B_{ij}}$ and $y_{ij} = b_{ij} \sqrt{a_{ij}}$ yields
\begin{align}\label{eq:B-RKUR}
    {\rm Var}{\bm(} \Theta(\tau) {\bm)} \geq \frac{R_\epsilon^2(\tau)}{\tau \sum_{i<j} b_{ij}^2 a_{ij}} \geq \frac{R_{\epsilon}^2(\tau)}{\tau b_{\rm max}^2 \dot{A}},
\end{align}
while choosing $x_{ij} = f_{ij} R_{F_{ij}}$, and $y_{ij} = f_{ij} \sqrt{a_{ij}}$ yields
\begin{align}
    {\rm Var}{\bm(} \Theta(\tau) {\bm)} \geq \frac{4R_\eta^2(\tau)}{\tau \sum_{i<j} f_{ij}^2 a_{ij}} \geq \frac{4R_\eta^2(\tau)}{\tau f_{\rm max}^2 \dot{A}},
\end{align}
where $b_{\rm max} = \max_{i,j} |b_{ij}|$ and $f_{\rm max} = \max_{i,j} |f_{ij}|$.
These inequalities show that the ratio of the responses to kinetic or entropic perturbations to fluctuations is bounded from above by the dynamical activity of the unperturbed system.

Similarly, with the choice $x_{ij} = b_{ij} R_{B_{ij}}$ and $y_{ij} = b_{ij} J_{ij}/\sqrt{a_{ij}}$, applying the Cauchy-Schwartz inequality to \eqref{eq:FRI_B2} leads to
\begin{align}
    {\rm Var}{\bm(} \Theta(\tau) {\bm)} \geq \frac{2R_{\epsilon}^2(\tau)}{\tau b_{\rm max}^2 \dot{\Sigma}_{\rm ps}} ~~~ {\rm for} ~ \Theta \in \mathcal{S}.
\end{align}
We refer to this relation as the response thermodynamic-kinetic uncertainty relation (R-TKUR), as it imposes an upper bound on the response that depends on both EP rate (a thermodynamic quantity) and dynamical activity (a kinetic quantity).
This can be seen clearly upon noting that the pseudo-EP rate satisfies the following Jensen's inequality $\dot{\Sigma}_{\rm ps} = \sum_{i<j} 2a_{ij} \phi^2\bm{(} (J_{ij}/2a_{ij}) \ln(W_{ij}\pi_j/W_{ji}\pi_i) \bm{)} \leq 2\dot{A} \phi^2(\dot{\Sigma}/2\dot{A})$ with a concave function $\phi(x) = x/\psi(x)$, where $\psi(x)$ is the inverse function of $x \tanh x$~\cite{Vo2022}.
One of the two contributions, $\dot \Sigma$ or $\dot{A}$, dominates in the limiting cases: for $x\ll 1$, $\phi^2(x) \approx x$, and for $x\gg 1$, $\phi^2(x) \approx 1$. Thus, the R-TKUR reduces to the finite-time version of R-TUR~\cite{ptaszynski2024dissipation},
\begin{align}\label{eq:R-TKUR}
    \frac{R_\epsilon^2(\tau)}{\text{Var}{\bm(} \Theta(\tau){\bm )}} \le  
    \frac{\tau b^2 _\text{max} \dot{\Sigma}}{2} ~~~ ~ {\rm for} ~ \Theta \in \mathcal{S} \;,
\end{align}
near equilibrium, where $\dot{\Sigma} / \dot{A} \ll 1$, and reduces to \eqref{eq:B-RKUR} far from equilibrium, where $\dot{\Sigma} / \dot{A} \gg 1$.
It is worth noting that (\ref{eq:B-RKUR}-\ref{eq:R-TKUR}) hold for all observation times $\tau$, generalizing the static response relations derived in the limit $\tau \rightarrow \infty$ in~\cite{ptaszynski2024dissipation,aslyamov2024nonequilibrium2} to dynamic response.

\begin{figure}
    \includegraphics[width=\columnwidth]{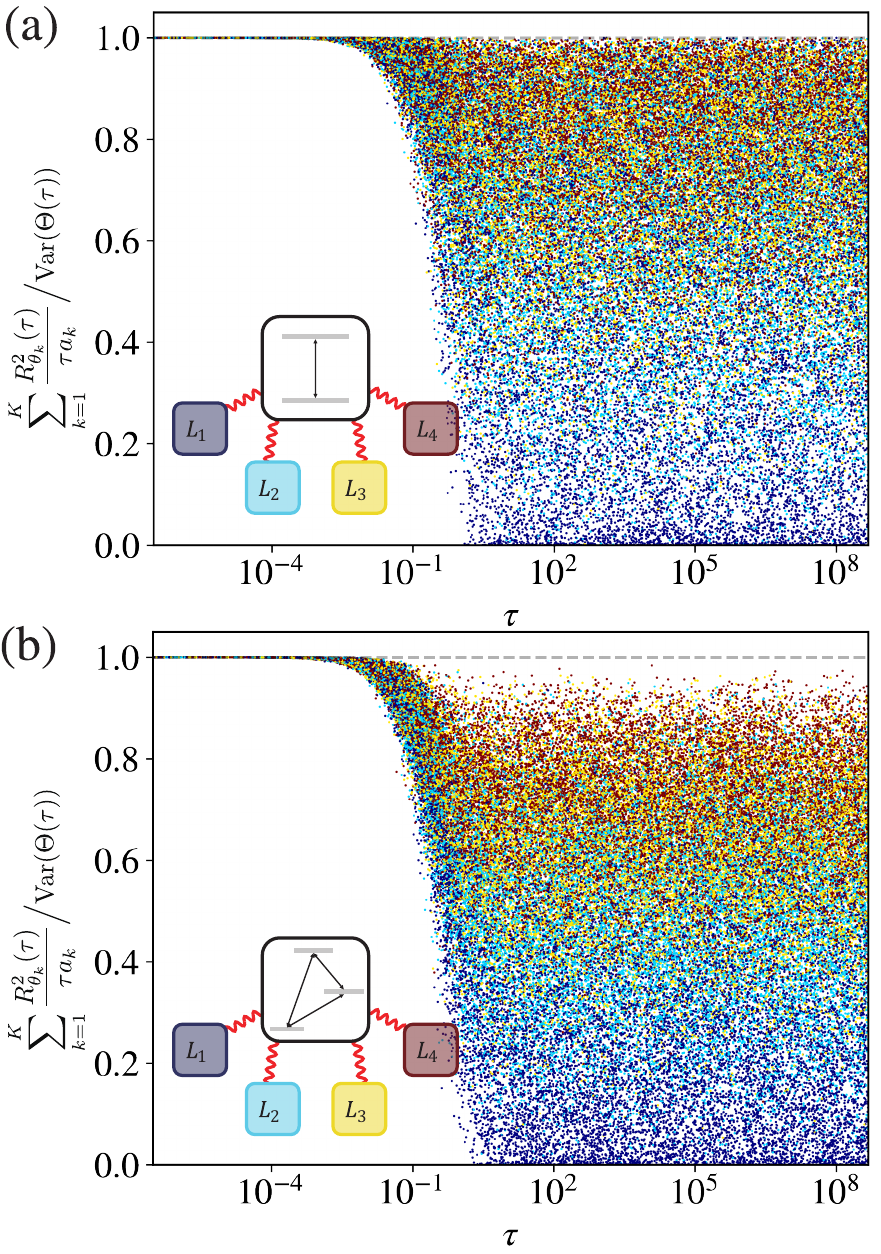}
    \caption{\label{fig:fig_FRI_quantum} Numerical verification of Eq.~\eqref{eq:FRI_Q} for (a) a 2-level system and (b) a 3-level system. The Hamiltonian is constructed as $(A + A^\dagger)/2$, where $A$ is a randomly generated matrix with the real and imaginary parts of each element independently and uniformly sampled from $[-1, 1]$. Jump operators are generated similarly without enforcing the Hermitian condition, with the number of jump operators randomly chosen between one and four. Different numbers of jump operators are represented by distinct colors: blue (1 channel), cyan (2 channels), yellow (3 channels), and dark red (4 channels), respectively. Weights $\Lambda_k$ are randomly sampled from $[-1,1]$, and the observation time $\tau$ is given by $\tau = e^x$, where $x$ is uniformly sampled from $[-15, 20]$.
}
    \end{figure}

\emph{Quantum generalization.}--The FRI can also be extended to Markovian open quantum systems, whose dynamics are governed by the Lindblad quantum master equation:
\begin{align}\label{eq:QME}
\dot{\rho}(t) =-i[H,\rho(t)]+\sum_{k=1} ^K \mathcal{D}[L_k ^{\theta_k} ]\rho (t)
\;,
\end{align}
where $\rho(t)$ is the density operator at time $t$, $H$ is the system Hamiltonian, $L_k^{\theta_k}$ are jump operators, $K$ denotes the number of channels, and $\mathcal{D}[L]\bullet \equiv L\bullet L^\dag-\{L^\dag L,\bullet\}/2$ is a superoperator acting on the operators to its right.
The Planck constant $\hbar$ is set to unity throughout.
In this section, the notations $[\bullet,\circ]$ and $\{ \bullet,\circ\}$ are reserved for the commutator and anti-commutator of two quantum operators $\bullet$ and $\circ$, respectively.
The jump operators are parameterized as 
\begin{align} \label{eq:quantum_parametrization}
    L_k ^{\theta_k} = e^{\theta_k /2} L_k
    \;,
\end{align} 
with a set of real-valued parameters $(\theta_1, \theta_2, \dots, \theta_K)$ that controls the magnitude of each jump operator.

As in the classical case, we consider the system in the steady state characterized by the density operator $\rho_\text{ss}$ for $t<0$, which is then slightly perturbed in its parameters $\theta_k$ at time $t=0$. The quantum Crame\'r-Rao bound generalizes the classical one as $\sum_{\alpha,\beta} R_{\theta_\alpha} (\tau) [\mathcal{I}^{-1}_Q (\tau)]_{\theta_\alpha \theta_\beta} R_{\theta_\beta}(\tau) \le \text{Var} {\bm(} \Theta(\tau) {\bm)}$, where $\mathcal{I}_Q(\tau)$ is the quantum Fisher information, defined as the maximum classical Fisher information over all positive operator-valued measures~\cite{helstrom1969quantum,paris2009quantum}. Similarly to the classical case, the quantum Fisher information matrix with respect to changes in $\theta_k$ becomes diagonal (see Appendix), which yields the quantum version of the FRI:
\begin{align} \label{eq:FRI_Q}
    \sum_{k=1} ^K \frac{R^2 _{\theta_k} (\tau)}{\tau a_k} \le \text{Var} {\bm (} \Theta(\tau) {\bm )}
    \;,
\end{align}
where $a_k = \text{tr}\bm{(}L_k^{\theta_k} \rho (L_k^{\theta_k})^\dagger \bm{)}$ is the traffic through the $k$-th channel.

The validity of the quantum FRI is examined in Fig.~\ref{fig:fig_FRI_quantum} for 2-level and 3-level systems, with randomly generated Hamiltonians and jump operators.
The local detailed balance condition is not applied to the jump operators.
The observables take the form $\Theta(\tau) =  \sum_{k=1} ^K \Lambda_k N_k (\tau)$, where $N_k(\tau)$ represents the total number of jumps through channel $k$ up to time $\tau$ and $\Lambda_k$ is the corresponding weight. Similarly to Fig.~\ref{fig:fig_FRI_B2}(a), the FRIs converge to equality in the limit $\tau\rightarrow 0$. We also note that the use of more jump operators results in a tighter inequality, although the precise relationship between the tightness and the number of jump operators requires further investigation. From the quantum FRI, a response uncertainty relation for open quantum systems can be derived straightforwardly:
\begin{align}
    \frac{R_\epsilon^2(\tau)}{\text{Var} {\bm (}\Theta(\tau) {\bm )}} \le \tau (\Delta \theta_\text{max})^2 \dot{A}
    \;,
\end{align}
where $\epsilon$ is the parameter that controls the magnitude of each jump operator via $\theta = \theta(\epsilon)$, $\Delta\theta_\text{max} = \max_k \{|d_\epsilon \theta_k|\}$, and $\dot{A} = \sum_{k} a_k$ is the dynamical activity.

\emph{Conclusion.}--In this Letter, we derive FRIs that apply to dynamic response to both kinetic and entropic perturbations, extending the previously established FRRs for static response.
These FRIs are valid for finite observation times and link the response function to fluctuations as well as to the thermodynamic and kinetic aspects of nonequilibrium systems.
Applicable across a wide range of observables, including current-like and state-dependent types, the FRIs demonstrate broader utility than existing FRRs.
Furthermore, we derive response uncertainty relations for dynamic response to kinetic and entropic perturbations, highlighting the interplay between entropy production and dynamical activity in constraining the system’s response.

The FRIs and response uncertainty relations are further extended to open quantum systems governed by the Lindblad quantum master equation. The resulting quantum FRI and response uncertainty relation involve only dynamical activity, representing the kinetic aspect of nonequilibrium systems.
It remains an open question whether similar relations can be found that incorporate EP, the thermodynamic aspect of nonequilibrium systems.
These findings provide a unified theoretical basis for exploring the response of nonequilibrium systems in both classical and quantum domains.

\emph{Acknowledgements.}--E.K and H.-M.C. equally contributed to this work. The authors acknowledge the Korea Institute for Advanced Study for providing computing resources (KIAS Center for Advanced Computation Linux Cluster System). This research was supported by NRF Grants No.~2017R1D1A1B06035497 (H.P.), No.~RS-2023-00278985 (E.K.), and individual KIAS Grants No.~PG064902 (J.S.L.), PG089402 (H.-M.C), and QP013601 (H.P.) at the Korea Institute for Advanced Study.

\bibliography{Response_TKUR}

\begin{thebibliography}{34}%
\makeatletter
\providecommand \@ifxundefined [1]{%
 \@ifx{#1\undefined}
}%
\providecommand \@ifnum [1]{%
 \ifnum #1\expandafter \@firstoftwo
 \else \expandafter \@secondoftwo
 \fi
}%
\providecommand \@ifx [1]{%
 \ifx #1\expandafter \@firstoftwo
 \else \expandafter \@secondoftwo
 \fi
}%
\providecommand \natexlab [1]{#1}%
\providecommand \enquote  [1]{``#1''}%
\providecommand \bibnamefont  [1]{#1}%
\providecommand \bibfnamefont [1]{#1}%
\providecommand \citenamefont [1]{#1}%
\providecommand \href@noop [0]{\@secondoftwo}%
\providecommand \href [0]{\begingroup \@sanitize@url \@href}%
\providecommand \@href[1]{\@@startlink{#1}\@@href}%
\providecommand \@@href[1]{\endgroup#1\@@endlink}%
\providecommand \@sanitize@url [0]{\catcode `\\12\catcode `\$12\catcode `\&12\catcode `\#12\catcode `\^12\catcode `\_12\catcode `\%12\relax}%
\providecommand \@@startlink[1]{}%
\providecommand \@@endlink[0]{}%
\providecommand \url  [0]{\begingroup\@sanitize@url \@url }%
\providecommand \@url [1]{\endgroup\@href {#1}{\urlprefix }}%
\providecommand \urlprefix  [0]{URL }%
\providecommand \Eprint [0]{\href }%
\providecommand \doibase [0]{http://dx.doi.org/}%
\providecommand \selectlanguage [0]{\@gobble}%
\providecommand \bibinfo  [0]{\@secondoftwo}%
\providecommand \bibfield  [0]{\@secondoftwo}%
\providecommand \translation [1]{[#1]}%
\providecommand \BibitemOpen [0]{}%
\providecommand \bibitemStop [0]{}%
\providecommand \bibitemNoStop [0]{.\EOS\space}%
\providecommand \EOS [0]{\spacefactor3000\relax}%
\providecommand \BibitemShut  [1]{\csname bibitem#1\endcsname}%
\let\auto@bib@innerbib\@empty
\bibitem [{\citenamefont {Chaikin}\ \emph {et~al.}(1995)\citenamefont {Chaikin}, \citenamefont {Lubensky},\ and\ \citenamefont {Witten}}]{chaikin1995principles}%
  \BibitemOpen
  \bibfield  {author} {\bibinfo {author} {\bibfnamefont {Paul~M}\ \bibnamefont {Chaikin}}, \bibinfo {author} {\bibfnamefont {Tom~C}\ \bibnamefont {Lubensky}}, \ and\ \bibinfo {author} {\bibfnamefont {Thomas~A}\ \bibnamefont {Witten}},\ }\href@noop {} {\emph {\bibinfo {title} {Principles of condensed matter physics}}},\ Vol.~\bibinfo {volume} {10}\ (\bibinfo  {publisher} {Cambridge university press Cambridge},\ \bibinfo {year} {1995})\BibitemShut {NoStop}%
\bibitem [{\citenamefont {Kubo}\ \emph {et~al.}(2012)\citenamefont {Kubo}, \citenamefont {Toda},\ and\ \citenamefont {Hashitsume}}]{kubo2012statistical}%
  \BibitemOpen
  \bibfield  {author} {\bibinfo {author} {\bibfnamefont {Ryogo}\ \bibnamefont {Kubo}}, \bibinfo {author} {\bibfnamefont {Morikazu}\ \bibnamefont {Toda}}, \ and\ \bibinfo {author} {\bibfnamefont {Natsuki}\ \bibnamefont {Hashitsume}},\ }\href@noop {} {\emph {\bibinfo {title} {Statistical physics II: nonequilibrium statistical mechanics}}},\ Vol.~\bibinfo {volume} {31}\ (\bibinfo  {publisher} {Springer Science \& Business Media},\ \bibinfo {year} {2012})\BibitemShut {NoStop}%
\bibitem [{\citenamefont {Mason}\ and\ \citenamefont {Weitz}(1995)}]{mason1995optical}%
  \BibitemOpen
  \bibfield  {author} {\bibinfo {author} {\bibfnamefont {Thomas~G}\ \bibnamefont {Mason}}\ and\ \bibinfo {author} {\bibfnamefont {David~A}\ \bibnamefont {Weitz}},\ }\bibfield  {title} {\enquote {\bibinfo {title} {Optical measurements of frequency-dependent linear viscoelastic moduli of complex fluids},}\ }\href@noop {} {\bibfield  {journal} {\bibinfo  {journal} {Physical Review Letters}\ }\textbf {\bibinfo {volume} {74}},\ \bibinfo {pages} {1250} (\bibinfo {year} {1995})}\BibitemShut {NoStop}%
\bibitem [{\citenamefont {Kubo}(1966)}]{kubo1966fluctuation}%
  \BibitemOpen
  \bibfield  {author} {\bibinfo {author} {\bibfnamefont {Ryogo}\ \bibnamefont {Kubo}},\ }\bibfield  {title} {\enquote {\bibinfo {title} {The fluctuation-dissipation theorem},}\ }\href@noop {} {\bibfield  {journal} {\bibinfo  {journal} {Reports on Progress in Physics}\ }\textbf {\bibinfo {volume} {29}},\ \bibinfo {pages} {255} (\bibinfo {year} {1966})}\BibitemShut {NoStop}%
\bibitem [{\citenamefont {Agarwal}(1972)}]{agarwal1972fluctuation}%
  \BibitemOpen
  \bibfield  {author} {\bibinfo {author} {\bibfnamefont {Girish~Saran}\ \bibnamefont {Agarwal}},\ }\bibfield  {title} {\enquote {\bibinfo {title} {Fluctuation-dissipation theorems for systems in non-thermal equilibrium and applications},}\ }\href@noop {} {\bibfield  {journal} {\bibinfo  {journal} {Zeitschrift f{\"u}r Physik A Hadrons and Nuclei}\ }\textbf {\bibinfo {volume} {252}},\ \bibinfo {pages} {25--38} (\bibinfo {year} {1972})}\BibitemShut {NoStop}%
\bibitem [{\citenamefont {Speck}\ and\ \citenamefont {Seifert}(2006)}]{speck2006restoring}%
  \BibitemOpen
  \bibfield  {author} {\bibinfo {author} {\bibfnamefont {Thomas}\ \bibnamefont {Speck}}\ and\ \bibinfo {author} {\bibfnamefont {Udo}\ \bibnamefont {Seifert}},\ }\bibfield  {title} {\enquote {\bibinfo {title} {Restoring a fluctuation-dissipation theorem in a nonequilibrium steady state},}\ }\href@noop {} {\bibfield  {journal} {\bibinfo  {journal} {Europhysics Letters}\ }\textbf {\bibinfo {volume} {74}},\ \bibinfo {pages} {391} (\bibinfo {year} {2006})}\BibitemShut {NoStop}%
\bibitem [{\citenamefont {Baiesi}\ \emph {et~al.}(2009)\citenamefont {Baiesi}, \citenamefont {Maes},\ and\ \citenamefont {Wynants}}]{baiesi2009fluctuations}%
  \BibitemOpen
  \bibfield  {author} {\bibinfo {author} {\bibfnamefont {Marco}\ \bibnamefont {Baiesi}}, \bibinfo {author} {\bibfnamefont {Christian}\ \bibnamefont {Maes}}, \ and\ \bibinfo {author} {\bibfnamefont {Bram}\ \bibnamefont {Wynants}},\ }\bibfield  {title} {\enquote {\bibinfo {title} {Fluctuations and response of nonequilibrium states},}\ }\href@noop {} {\bibfield  {journal} {\bibinfo  {journal} {Physical Review Letters}\ }\textbf {\bibinfo {volume} {103}},\ \bibinfo {pages} {010602} (\bibinfo {year} {2009})}\BibitemShut {NoStop}%
\bibitem [{\citenamefont {Seifert}\ and\ \citenamefont {Speck}(2010)}]{seifert2010fluctuation}%
  \BibitemOpen
  \bibfield  {author} {\bibinfo {author} {\bibfnamefont {Udo}\ \bibnamefont {Seifert}}\ and\ \bibinfo {author} {\bibfnamefont {Thomas}\ \bibnamefont {Speck}},\ }\bibfield  {title} {\enquote {\bibinfo {title} {Fluctuation-dissipation theorem in nonequilibrium steady states},}\ }\href@noop {} {\bibfield  {journal} {\bibinfo  {journal} {Europhysics Letters}\ }\textbf {\bibinfo {volume} {89}},\ \bibinfo {pages} {10007} (\bibinfo {year} {2010})}\BibitemShut {NoStop}%
\bibitem [{\citenamefont {Altaner}\ \emph {et~al.}(2016)\citenamefont {Altaner}, \citenamefont {Polettini},\ and\ \citenamefont {Esposito}}]{altaner2016fluctuation}%
  \BibitemOpen
  \bibfield  {author} {\bibinfo {author} {\bibfnamefont {Bernhard}\ \bibnamefont {Altaner}}, \bibinfo {author} {\bibfnamefont {Matteo}\ \bibnamefont {Polettini}}, \ and\ \bibinfo {author} {\bibfnamefont {Massimiliano}\ \bibnamefont {Esposito}},\ }\bibfield  {title} {\enquote {\bibinfo {title} {Fluctuation-dissipation relations far from equilibrium},}\ }\href@noop {} {\bibfield  {journal} {\bibinfo  {journal} {Physical Review Letters}\ }\textbf {\bibinfo {volume} {117}},\ \bibinfo {pages} {180601} (\bibinfo {year} {2016})}\BibitemShut {NoStop}%
\bibitem [{\citenamefont {Owen}\ \emph {et~al.}(2020)\citenamefont {Owen}, \citenamefont {Gingrich},\ and\ \citenamefont {Horowitz}}]{owen2020universal}%
  \BibitemOpen
  \bibfield  {author} {\bibinfo {author} {\bibfnamefont {Jeremy~A}\ \bibnamefont {Owen}}, \bibinfo {author} {\bibfnamefont {Todd~R}\ \bibnamefont {Gingrich}}, \ and\ \bibinfo {author} {\bibfnamefont {Jordan~M}\ \bibnamefont {Horowitz}},\ }\bibfield  {title} {\enquote {\bibinfo {title} {Universal thermodynamic bounds on nonequilibrium response with biochemical applications},}\ }\href@noop {} {\bibfield  {journal} {\bibinfo  {journal} {Physical Review X}\ }\textbf {\bibinfo {volume} {10}},\ \bibinfo {pages} {011066} (\bibinfo {year} {2020})}\BibitemShut {NoStop}%
\bibitem [{\citenamefont {Aslyamov}\ and\ \citenamefont {Esposito}(2024{\natexlab{a}})}]{aslyamov2024nonequilibrium}%
  \BibitemOpen
  \bibfield  {author} {\bibinfo {author} {\bibfnamefont {Timur}\ \bibnamefont {Aslyamov}}\ and\ \bibinfo {author} {\bibfnamefont {Massimiliano}\ \bibnamefont {Esposito}},\ }\bibfield  {title} {\enquote {\bibinfo {title} {Nonequilibrium response for markov jump processes: Exact results and tight bounds},}\ }\href@noop {} {\bibfield  {journal} {\bibinfo  {journal} {Physical Review Letters}\ }\textbf {\bibinfo {volume} {132}},\ \bibinfo {pages} {037101} (\bibinfo {year} {2024}{\natexlab{a}})}\BibitemShut {NoStop}%
\bibitem [{\citenamefont {Dechant}\ and\ \citenamefont {Sasa}(2020)}]{dechant2020fluctuation}%
  \BibitemOpen
  \bibfield  {author} {\bibinfo {author} {\bibfnamefont {Andreas}\ \bibnamefont {Dechant}}\ and\ \bibinfo {author} {\bibfnamefont {Shin-ichi}\ \bibnamefont {Sasa}},\ }\bibfield  {title} {\enquote {\bibinfo {title} {Fluctuation--response inequality out of equilibrium},}\ }\href@noop {} {\bibfield  {journal} {\bibinfo  {journal} {Proceedings of the National Academy of Sciences}\ }\textbf {\bibinfo {volume} {117}},\ \bibinfo {pages} {6430--6436} (\bibinfo {year} {2020})}\BibitemShut {NoStop}%
\bibitem [{\citenamefont {Gao}\ \emph {et~al.}(2022)\citenamefont {Gao}, \citenamefont {Chun},\ and\ \citenamefont {Horowitz}}]{gao2022thermodynamic}%
  \BibitemOpen
  \bibfield  {author} {\bibinfo {author} {\bibfnamefont {Qi}~\bibnamefont {Gao}}, \bibinfo {author} {\bibfnamefont {Hyun-Myung}\ \bibnamefont {Chun}}, \ and\ \bibinfo {author} {\bibfnamefont {Jordan~M}\ \bibnamefont {Horowitz}},\ }\bibfield  {title} {\enquote {\bibinfo {title} {Thermodynamic constraints on the nonequilibrium response of one-dimensional diffusions},}\ }\href@noop {} {\bibfield  {journal} {\bibinfo  {journal} {Physical Review E}\ }\textbf {\bibinfo {volume} {105}},\ \bibinfo {pages} {L012102} (\bibinfo {year} {2022})}\BibitemShut {NoStop}%
\bibitem [{\citenamefont {Chun}\ and\ \citenamefont {Horowitz}(2023)}]{chun2023trade}%
  \BibitemOpen
  \bibfield  {author} {\bibinfo {author} {\bibfnamefont {Hyun-Myung}\ \bibnamefont {Chun}}\ and\ \bibinfo {author} {\bibfnamefont {Jordan~M}\ \bibnamefont {Horowitz}},\ }\bibfield  {title} {\enquote {\bibinfo {title} {Trade-offs between number fluctuations and response in nonequilibrium chemical reaction networks},}\ }\href@noop {} {\bibfield  {journal} {\bibinfo  {journal} {The Journal of Chemical Physics}\ }\textbf {\bibinfo {volume} {158}} (\bibinfo {year} {2023})}\BibitemShut {NoStop}%
\bibitem [{\citenamefont {Fernandes~Martins}\ and\ \citenamefont {Horowitz}(2023)}]{fernandes2023topologically}%
  \BibitemOpen
  \bibfield  {author} {\bibinfo {author} {\bibfnamefont {Gabriela}\ \bibnamefont {Fernandes~Martins}}\ and\ \bibinfo {author} {\bibfnamefont {Jordan~M}\ \bibnamefont {Horowitz}},\ }\bibfield  {title} {\enquote {\bibinfo {title} {Topologically constrained fluctuations and thermodynamics regulate nonequilibrium response},}\ }\href@noop {} {\bibfield  {journal} {\bibinfo  {journal} {Physical Review E}\ }\textbf {\bibinfo {volume} {108}},\ \bibinfo {pages} {044113} (\bibinfo {year} {2023})}\BibitemShut {NoStop}%
\bibitem [{\citenamefont {Gao}\ \emph {et~al.}(2024)\citenamefont {Gao}, \citenamefont {Chun},\ and\ \citenamefont {Horowitz}}]{gao2024thermodynamic}%
  \BibitemOpen
  \bibfield  {author} {\bibinfo {author} {\bibfnamefont {Qi}~\bibnamefont {Gao}}, \bibinfo {author} {\bibfnamefont {Hyun-Myung}\ \bibnamefont {Chun}}, \ and\ \bibinfo {author} {\bibfnamefont {Jordan~M}\ \bibnamefont {Horowitz}},\ }\bibfield  {title} {\enquote {\bibinfo {title} {Thermodynamic constraints on kinetic perturbations of homogeneous driven diffusions},}\ }\href@noop {} {\bibfield  {journal} {\bibinfo  {journal} {Europhysics Letters}\ }\textbf {\bibinfo {volume} {146}},\ \bibinfo {pages} {31001} (\bibinfo {year} {2024})}\BibitemShut {NoStop}%
\bibitem [{\citenamefont {Aslyamov}\ and\ \citenamefont {Esposito}(2024{\natexlab{b}})}]{aslyamov2024general}%
  \BibitemOpen
  \bibfield  {author} {\bibinfo {author} {\bibfnamefont {Timur}\ \bibnamefont {Aslyamov}}\ and\ \bibinfo {author} {\bibfnamefont {Massimiliano}\ \bibnamefont {Esposito}},\ }\bibfield  {title} {\enquote {\bibinfo {title} {General theory of static response for markov jump processes},}\ }\href@noop {} {\bibfield  {journal} {\bibinfo  {journal} {Physical Review Letters}\ }\textbf {\bibinfo {volume} {133}},\ \bibinfo {pages} {107103} (\bibinfo {year} {2024}{\natexlab{b}})}\BibitemShut {NoStop}%
\bibitem [{\citenamefont {Ptaszynski}\ \emph {et~al.}(2024)\citenamefont {Ptaszynski}, \citenamefont {Aslyamov},\ and\ \citenamefont {Esposito}}]{ptaszynski2024dissipation}%
  \BibitemOpen
  \bibfield  {author} {\bibinfo {author} {\bibfnamefont {Krzysztof}\ \bibnamefont {Ptaszynski}}, \bibinfo {author} {\bibfnamefont {Timur}\ \bibnamefont {Aslyamov}}, \ and\ \bibinfo {author} {\bibfnamefont {Massimiliano}\ \bibnamefont {Esposito}},\ }\bibfield  {title} {\enquote {\bibinfo {title} {Dissipation bounds precision of current response to kinetic perturbations},}\ }\href@noop {} {\bibfield  {journal} {\bibinfo  {journal} {arXiv preprint arXiv:2406.08361}\ } (\bibinfo {year} {2024})}\BibitemShut {NoStop}%
\bibitem [{\citenamefont {Aslyamov}\ \emph {et~al.}(2024)\citenamefont {Aslyamov}, \citenamefont {Ptaszy{\'n}ski},\ and\ \citenamefont {Esposito}}]{aslyamov2024nonequilibrium2}%
  \BibitemOpen
  \bibfield  {author} {\bibinfo {author} {\bibfnamefont {Timur}\ \bibnamefont {Aslyamov}}, \bibinfo {author} {\bibfnamefont {Krzysztof}\ \bibnamefont {Ptaszy{\'n}ski}}, \ and\ \bibinfo {author} {\bibfnamefont {Massimiliano}\ \bibnamefont {Esposito}},\ }\bibfield  {title} {\enquote {\bibinfo {title} {Nonequilibrium fluctuation-response relations: From identities to bounds},}\ }\href@noop {} {\bibfield  {journal} {\bibinfo  {journal} {arXiv preprint arXiv:2410.17140}\ } (\bibinfo {year} {2024})}\BibitemShut {NoStop}%
\bibitem [{\citenamefont {Dechant}(2018)}]{dechant2018multidimensional}%
  \BibitemOpen
  \bibfield  {author} {\bibinfo {author} {\bibfnamefont {Andreas}\ \bibnamefont {Dechant}},\ }\bibfield  {title} {\enquote {\bibinfo {title} {Multidimensional thermodynamic uncertainty relations},}\ }\href@noop {} {\bibfield  {journal} {\bibinfo  {journal} {Journal of Physics A: Mathematical and Theoretical}\ }\textbf {\bibinfo {volume} {52}},\ \bibinfo {pages} {035001} (\bibinfo {year} {2018})}\BibitemShut {NoStop}%
\bibitem [{\citenamefont {Hasegawa}\ and\ \citenamefont {Van~Vu}(2019)}]{hasegawa2019uncertainty}%
  \BibitemOpen
  \bibfield  {author} {\bibinfo {author} {\bibfnamefont {Yoshihiko}\ \bibnamefont {Hasegawa}}\ and\ \bibinfo {author} {\bibfnamefont {Tan}\ \bibnamefont {Van~Vu}},\ }\bibfield  {title} {\enquote {\bibinfo {title} {Uncertainty relations in stochastic processes: An information inequality approach},}\ }\href@noop {} {\bibfield  {journal} {\bibinfo  {journal} {Physical Review E}\ }\textbf {\bibinfo {volume} {99}},\ \bibinfo {pages} {062126} (\bibinfo {year} {2019})}\BibitemShut {NoStop}%
\bibitem [{\citenamefont {Lee}\ \emph {et~al.}(2019)\citenamefont {Lee}, \citenamefont {Park},\ and\ \citenamefont {Park}}]{lee2019thermodynamic}%
  \BibitemOpen
  \bibfield  {author} {\bibinfo {author} {\bibfnamefont {Jae~Sung}\ \bibnamefont {Lee}}, \bibinfo {author} {\bibfnamefont {Jong-Min}\ \bibnamefont {Park}}, \ and\ \bibinfo {author} {\bibfnamefont {Hyunggyu}\ \bibnamefont {Park}},\ }\bibfield  {title} {\enquote {\bibinfo {title} {Thermodynamic uncertainty relation for underdamped langevin systems driven by a velocity-dependent force},}\ }\href@noop {} {\bibfield  {journal} {\bibinfo  {journal} {Physical Review E}\ }\textbf {\bibinfo {volume} {100}},\ \bibinfo {pages} {062132} (\bibinfo {year} {2019})}\BibitemShut {NoStop}%
\bibitem [{\citenamefont {Lee}\ \emph {et~al.}(2021)\citenamefont {Lee}, \citenamefont {Park},\ and\ \citenamefont {Park}}]{lee2021universal}%
  \BibitemOpen
  \bibfield  {author} {\bibinfo {author} {\bibfnamefont {Jae~Sung}\ \bibnamefont {Lee}}, \bibinfo {author} {\bibfnamefont {Jong-Min}\ \bibnamefont {Park}}, \ and\ \bibinfo {author} {\bibfnamefont {Hyunggyu}\ \bibnamefont {Park}},\ }\bibfield  {title} {\enquote {\bibinfo {title} {Universal form of thermodynamic uncertainty relation for langevin dynamics},}\ }\href@noop {} {\bibfield  {journal} {\bibinfo  {journal} {Physical Review E}\ }\textbf {\bibinfo {volume} {104}},\ \bibinfo {pages} {L052102} (\bibinfo {year} {2021})}\BibitemShut {NoStop}%
\bibitem [{\citenamefont {Vo}\ \emph {et~al.}(2022)\citenamefont {Vo}, \citenamefont {Van~Vu},\ and\ \citenamefont {Hasegawa}}]{Vo2022}%
  \BibitemOpen
  \bibfield  {author} {\bibinfo {author} {\bibfnamefont {Van~Tuan}\ \bibnamefont {Vo}}, \bibinfo {author} {\bibfnamefont {Tan}\ \bibnamefont {Van~Vu}}, \ and\ \bibinfo {author} {\bibfnamefont {Yoshihiko}\ \bibnamefont {Hasegawa}},\ }\bibfield  {title} {\enquote {\bibinfo {title} {Unified thermodynamic–kinetic uncertainty relation},}\ }\href {\doibase 10.1088/1751-8121/ac9099} {\bibfield  {journal} {\bibinfo  {journal} {Journal of Physics A: Mathematical and Theoretical}\ }\textbf {\bibinfo {volume} {55}},\ \bibinfo {pages} {405004} (\bibinfo {year} {2022})}\BibitemShut {NoStop}%
\bibitem [{\citenamefont {Seifert}(2012)}]{seifert2012stochastic}%
  \BibitemOpen
  \bibfield  {author} {\bibinfo {author} {\bibfnamefont {Udo}\ \bibnamefont {Seifert}},\ }\bibfield  {title} {\enquote {\bibinfo {title} {Stochastic thermodynamics, fluctuation theorems and molecular machines},}\ }\href@noop {} {\bibfield  {journal} {\bibinfo  {journal} {Reports on progress in physics}\ }\textbf {\bibinfo {volume} {75}},\ \bibinfo {pages} {126001} (\bibinfo {year} {2012})}\BibitemShut {NoStop}%
\bibitem [{\citenamefont {Maes}(2021)}]{maes2021local}%
  \BibitemOpen
  \bibfield  {author} {\bibinfo {author} {\bibfnamefont {Christian}\ \bibnamefont {Maes}},\ }\bibfield  {title} {\enquote {\bibinfo {title} {Local detailed balance},}\ }\href@noop {} {\bibfield  {journal} {\bibinfo  {journal} {SciPost Physics Lecture Notes}\ ,\ \bibinfo {pages} {32}} (\bibinfo {year} {2021})}\BibitemShut {NoStop}%
\bibitem [{\citenamefont {Shiraishi}(2021)}]{Shiraishi2021}%
  \BibitemOpen
  \bibfield  {author} {\bibinfo {author} {\bibfnamefont {Naoto}\ \bibnamefont {Shiraishi}},\ }\bibfield  {title} {\enquote {\bibinfo {title} {{Optimal Thermodynamic Uncertainty Relation in Markov Jump Processes}},}\ }\href {\doibase 10.1007/s10955-021-02829-8} {\bibfield  {journal} {\bibinfo  {journal} {Journal of Statistical Physics}\ }\textbf {\bibinfo {volume} {185}},\ \bibinfo {pages} {19} (\bibinfo {year} {2021})}\BibitemShut {NoStop}%
\bibitem [{\citenamefont {Kwon}\ \emph {et~al.}(2024)\citenamefont {Kwon}, \citenamefont {Park}, \citenamefont {Lee},\ and\ \citenamefont {Baek}}]{kwon2024unified}%
  \BibitemOpen
  \bibfield  {author} {\bibinfo {author} {\bibfnamefont {Euijoon}\ \bibnamefont {Kwon}}, \bibinfo {author} {\bibfnamefont {Jong-Min}\ \bibnamefont {Park}}, \bibinfo {author} {\bibfnamefont {Jae~Sung}\ \bibnamefont {Lee}}, \ and\ \bibinfo {author} {\bibfnamefont {Yongjoo}\ \bibnamefont {Baek}},\ }\bibfield  {title} {\enquote {\bibinfo {title} {Unified hierarchical relationship between thermodynamic tradeoff relations},}\ }\href@noop {} {\bibfield  {journal} {\bibinfo  {journal} {Physical Review E}\ }\textbf {\bibinfo {volume} {110}},\ \bibinfo {pages} {044131} (\bibinfo {year} {2024})}\BibitemShut {NoStop}%
\bibitem [{\citenamefont {Maes}(2020)}]{maes2020frenesy}%
  \BibitemOpen
  \bibfield  {author} {\bibinfo {author} {\bibfnamefont {Christian}\ \bibnamefont {Maes}},\ }\bibfield  {title} {\enquote {\bibinfo {title} {Frenesy: Time-symmetric dynamical activity in nonequilibria},}\ }\href@noop {} {\bibfield  {journal} {\bibinfo  {journal} {Physics Reports}\ }\textbf {\bibinfo {volume} {850}},\ \bibinfo {pages} {1--33} (\bibinfo {year} {2020})}\BibitemShut {NoStop}%
\bibitem [{\citenamefont {Kay}(1993)}]{kay1993fundamentals}%
  \BibitemOpen
  \bibfield  {author} {\bibinfo {author} {\bibfnamefont {Steven~M}\ \bibnamefont {Kay}},\ }\href@noop {} {\emph {\bibinfo {title} {Fundamentals of Statistical Signal Processing: Estimation Theory}}}\ (\bibinfo  {publisher} {Prentice Hall},\ \bibinfo {year} {1993})\BibitemShut {NoStop}%
\bibitem [{\citenamefont {Helstrom}(1969)}]{helstrom1969quantum}%
  \BibitemOpen
  \bibfield  {author} {\bibinfo {author} {\bibfnamefont {Carl~W}\ \bibnamefont {Helstrom}},\ }\bibfield  {title} {\enquote {\bibinfo {title} {Quantum detection and estimation theory},}\ }\href@noop {} {\bibfield  {journal} {\bibinfo  {journal} {Journal of Statistical Physics}\ }\textbf {\bibinfo {volume} {1}},\ \bibinfo {pages} {231--252} (\bibinfo {year} {1969})}\BibitemShut {NoStop}%
\bibitem [{\citenamefont {Paris}(2009)}]{paris2009quantum}%
  \BibitemOpen
  \bibfield  {author} {\bibinfo {author} {\bibfnamefont {Matteo G.~A.}\ \bibnamefont {Paris}},\ }\bibfield  {title} {\enquote {\bibinfo {title} {Quantum estimation for quantum technology},}\ }\href@noop {} {\bibfield  {journal} {\bibinfo  {journal} {International Journal of Quantum Information}\ }\textbf {\bibinfo {volume} {7}},\ \bibinfo {pages} {125--137} (\bibinfo {year} {2009})}\BibitemShut {NoStop}%
\bibitem [{\citenamefont {Gammelmark}\ and\ \citenamefont {M\o{}lmer}(2014)}]{gammelmark2014fisher}%
  \BibitemOpen
  \bibfield  {author} {\bibinfo {author} {\bibfnamefont {S\o{}ren}\ \bibnamefont {Gammelmark}}\ and\ \bibinfo {author} {\bibfnamefont {Klaus}\ \bibnamefont {M\o{}lmer}},\ }\bibfield  {title} {\enquote {\bibinfo {title} {Fisher information and the quantum cram\'er-rao sensitivity limit of continuous measurements},}\ }\href {\doibase 10.1103/PhysRevLett.112.170401} {\bibfield  {journal} {\bibinfo  {journal} {Physical Review Letters}\ }\textbf {\bibinfo {volume} {112}},\ \bibinfo {pages} {170401} (\bibinfo {year} {2014})}\BibitemShut {NoStop}%
\bibitem [{\citenamefont {Gardiner}(1985)}]{gardiner1985handbook}%
  \BibitemOpen
  \bibfield  {author} {\bibinfo {author} {\bibfnamefont {Crispin~W}\ \bibnamefont {Gardiner}},\ }\bibfield  {title} {\enquote {\bibinfo {title} {Handbook of stochastic methods for physics, chemistry and the natural sciences},}\ }\href@noop {} {\bibfield  {journal} {\bibinfo  {journal} {Springer series in synergetics}\ } (\bibinfo {year} {1985})}\BibitemShut {NoStop}%
\end{thebibliography}%

\section{Appendix}

\subsection{Derivation of fluctuation-response inequalities} \label{App:FRI_derivation}
Here, we derive the classical and quantum FRIs presented in \eqref{eq:FRI_B}, \eqref{eq:FRI_F}, and \eqref{eq:FRI_Q}.
In classical Markov jump processes, the probability density of observing the trajectory $\Gamma_\tau = \{ s(t) \}_{t=0}^\tau$ up to time $\tau$ is given by $\mathcal{P}[\Gamma_\tau] = \pi_{s(0)} e^{-\mathcal{A}[\Gamma_\tau]}$ with
\begin{align}
    \mathcal{A}[\Gamma_\tau] = \int_0^\tau dt \sum_{i\neq j} \left( \eta_j(t) W_{ij} - \dot{N}_{ij}(t)\ln W_{ij} \right) \;.
\end{align}
The elements of the Fisher information matrix associated with the perturbation parameters $(\theta_1,\cdots,\theta_K)$ are obtained by directly differentiating the path probability, yielding
\begin{equation} \label{eq:FIM_element}
\begin{aligned}
    & \mathcal{I}_{\theta_\alpha \theta_\beta}(\tau) 
    = -\langle \partial_{\theta_\alpha} \partial_{\theta_\beta} \ln \mathcal{P}[\Gamma_\tau] \rangle_0
    = \langle \partial_{\theta_\alpha} \partial_{\theta_\beta} \mathcal{A}[\Gamma_\tau] \rangle_0 \\
    &= \int_0^\tau dt \sum_{i\neq j} (\pi_j \partial_{\theta_\alpha} \partial_{\theta_\beta} W_{ij} - \pi_j W_{ij} \partial_{\theta_\alpha} \partial_{\theta_\beta} \ln W_{ij}) \\
    & = \tau \sum_{i\neq j} W_{ij} \pi_j \left( \partial_{\theta_\alpha} \ln W_{ij} \right) \left( \partial_{\theta_\beta} \ln W_{ij} \right)
    \;.
\end{aligned}
\end{equation}
Note that the initial distribution does not contribute to the Fisher information matrix since the perturbation considered does not alter the initial condition.
When the perturbation parameters are chosen as either the symmetric parameters $B_{ij}$ or the anti-symmetric parameters $F_{ij}$, the following relations hold:
\begin{align} \label{eqA:B_F_Wij_relations}
    &\partial_{B_{i'j'}} \ln W_{ij} = \delta_{ii'} \delta_{jj'} + \delta_{ij'}\delta_{i'j}\;, \nonumber \\
    &\partial_{F_{i'j'}} \ln W_{ij} = \frac{1}{2}(\delta_{ii'} \delta_{jj'} - \delta_{ij'}\delta_{i'j})\;.
\end{align}
Substituting Eq.~\eqref{eqA:B_F_Wij_relations} into Eq.~\eqref{eq:FIM_element}, we find that the Fisher information matrix becomes diagonal, with elements
\begin{align}
    &\mathcal{I}_{B_{ij}B_{i'j'}}(\tau) = \tau \delta_{ii'}\delta_{jj'} a_{ij}\;, \nonumber \\
    &\mathcal{I}_{F_{ij}F_{i'j'}}(\tau) = \frac{1}{4} \tau \delta_{ii'}\delta_{jj'} a_{ij}
    \;.
\end{align}
Plugging the Fisher information matrix into the Cram\'er-Rao bound, $\sum_{\alpha\beta} R_{\theta_\alpha}(\tau)[\mathcal{I}^{-1}(\tau)]_{\theta_\alpha\theta_\beta} R_{\theta_\beta}$, with $\theta_\alpha$ being either $B_{ij}$ or $F_{ij}$, we arrive at the classical FRIs, \eqref{eq:FRI_B} and \eqref{eq:FRI_F}.

For open quantum systems described by the Lindblad quantum master equation~\eqref{eq:QME}, the continuous measurement framework allows the quantum Fisher information matrix to be expressed in terms of the solution of the generalized Lindblad equation $\dot{\rho}(\tau) = \mathcal{L}_{\theta^1 \theta^2} \rho(\tau)$~\cite{gammelmark2014fisher}, where the superoperator $\mathcal{L}_{\theta^1 \theta^2}$ is given as
\begin{equation} \label{eqA:tilted_Lindblad}
\begin{aligned}
    \mathcal{L}_{\theta^1 \theta^2} \bullet 
    & = -i[H, \bullet] 
    + \sum_{k=1} ^K L_k ^{\theta_k ^1} \bullet (L_k ^{\theta_k ^2})^\dagger
    \\ & \quad - \frac{1}{2} \sum_{k=1} ^K \left[ (L_k ^{\theta_k ^1})^\dagger L_k ^{\theta_k ^1} \bullet + \bullet (L_k ^{\theta_k ^2})^\dagger L_k ^{\theta_k ^2}  \right]
    \;.
\end{aligned}
\end{equation}
The unperturbed dynamics are restored by setting $\theta^1 = \theta^2 = \theta$, whose superoperator is denoted by $\mathcal{L}$ hereafter.
The initial condition is $\rho(0) = \rho_{\rm ss}$, where $\rho_{\rm ss}$ is the steady-state density operator of the unperturbed dynamics, satisfying $\mathcal{L} \rho_{\rm ss} = 0$.
Denoting the solution of Eq.~\eqref{eqA:tilted_Lindblad} as $\rho_{\theta^1\theta^2}$, the elements of the quantum Fisher information matrix are expressed as~\cite{gammelmark2014fisher}
\begin{align} \label{eq:FIM_Q_element}
    [\mathcal{I}_Q(\tau)]_{\theta_\alpha \theta_\beta} 
    &= 4 \partial_\alpha^1 \partial_\beta^2 \ln( {\rm tr}[ \rho_{\theta^1\theta^2}(\tau)])_{\theta^1=\theta^2 = \theta} \\ \nonumber
    &= 4\{\partial_\alpha ^1 \partial_\beta ^2 \text{tr}[\rho_{\theta^1 \theta^2}(\tau)] 
    \\ \nonumber & \quad\quad   - \partial_\alpha ^1 \text{tr}[\rho_{\theta^1 \theta^2}(\tau)] \partial_\beta ^2 \text{tr}[\rho_{\theta^1 \theta^2}(\tau)] \}_{\theta^1=\theta^2=\theta}
    \;,
\end{align}
where $\theta^1$ and $\theta^2$ are $K$-dimensional vectors, and $\partial_{\alpha(\beta)} ^{1(2)}$ denotes the derivative with respect to the $\alpha$-th ($\beta$-th) component of $\theta^1 (\theta ^2)$.

Using the parameterization~\eqref{eq:quantum_parametrization}, we can demonstrate that 
\begin{align} \label{eqA:traceless_map}
    &(\partial_{\alpha}^{1} \mathcal{L} )_{\theta^1 = \theta^2 = \theta} \bullet = \frac{1}{2} {\bm(} L_\alpha^{\theta_\alpha} \bullet (L_\alpha^{\theta_\alpha})^\dagger - (L_\alpha^{\theta_\alpha})^\dagger L_\alpha^{\theta_\alpha} \bullet {\bm)} \; \nonumber \\
    &(\partial_{\beta}^{2} \mathcal{L} )_{\theta^1 = \theta^2 = \theta} \bullet = \frac{1}{2} {\bm (} L_\beta^{\theta_\beta} \bullet (L_\beta^{\theta_\beta})^\dagger - \bullet (L_\beta^{\theta_\beta})^\dagger L_\beta^{\theta_\beta} {\bm )} \;.
\end{align}
are traceless maps. From Eq.~\eqref{eqA:traceless_map}, we can show that the second term in Eq.~\eqref{eq:FIM_Q_element} vanishes because
\begin{align}
    &\partial_{\alpha(\beta)} ^{1(2)} \text{tr}[\rho_{\theta^1 \theta^2}(\tau)]_{\theta^1 = \theta^2 = \theta} 
    \\ \nonumber &= \int_0 ^\tau dt ~ \text{tr}\left[e^{\mathcal{L}(\tau-t)}(\partial_{\alpha(\beta)} ^{1(2)} \mathcal{L})_{\theta^1 = \theta^2 = \theta} e^{\mathcal{L}t} \rho_\text{ss} \right]
    \\ \nonumber &= \int_0 ^\tau dt ~ \text{tr}\left[(\partial_{\alpha(\beta)} ^{1(2)} \mathcal{L})_{\theta^1 = \theta^2 = \theta} \rho_\text{ss} \right]
    = 0
    \;,
\end{align}
where we use the fact that $e^{\mathcal{L}(\tau-t)}$ is a trace-preserving map and that $e^{\mathcal{L}t}\rho_{\rm ss} = \rho_{\rm ss}$ in the second equality.

The first term in Eq.~\eqref{eq:FIM_Q_element} is evaluated as
\begin{align} \label{eqA:first_term_calc}
    &\partial_\alpha ^1 \partial_\beta ^2 \text{tr}[\rho_{\theta^1 \theta^2}(\tau)]_{\theta^1 = \theta^2 = \theta} \\ 
    &= \int_0 ^\tau dt ~ \text{tr}\left[e^{\mathcal{L}(\tau-t)}(\partial_{\alpha}^{1} \partial_{\beta} ^{2} \mathcal{L})_{\theta^1 = \theta^2 = \theta} e^{\mathcal{L}t} \rho_\text{ss} \right] \nonumber \\
    &+ \int_0 ^\tau dt ~ \text{tr}\left[ ( \partial_{\alpha} ^{1} e^{\mathcal{L}(\tau-t)} )_{\theta^1 = \theta^2 = \theta}(\partial_{\beta} ^{2} \mathcal{L})_{\theta^1 = \theta^2 = \theta} e^{\mathcal{L}t} \rho_\text{ss} \right] \nonumber \\
    &+ \int_0 ^\tau dt ~ \text{tr}\left[e^{\mathcal{L}(\tau-t)}(\partial_{\beta} ^{2} \mathcal{L})_{\theta^1 = \theta^2 = \theta} (\partial_{\alpha} ^{1} e^{\mathcal{L}t} )_{\theta^1 = \theta^2 = \theta} \rho_\text{ss}  \right] 
    \;. \nonumber
\end{align}
The third integral in Eq.~\eqref{eqA:first_term_calc} vanishes as $e^{\mathcal L (\tau -t)}$ is trace-preserving and $(\partial_{\beta} ^{2} \mathcal{L})_{\theta^1 = \theta^2 = \theta}$ is a traceless map. Using the relation 
\begin{align}
    \partial_{\alpha} ^{1} e^{\mathcal{L}(\tau-t)} = \int_0^{\tau - t} dt' e^{\mathcal{L}(\tau-t-t')}(\partial_{\alpha}^{1}  \mathcal{L}) e^{\mathcal{L}t'},   
\end{align}
we can show that the second integral in Eq.~\eqref{eqA:first_term_calc} also vanishes. Thus, Eq.~\eqref{eq:FIM_Q_element} simplifies to
\begin{align}
    &\partial_\alpha ^1 \partial_\beta ^2 \text{tr}[\rho_{\theta^1 \theta^2}(\tau)]_{\theta^1 = \theta^2 = \theta} 
    \\ \nonumber &= \int_0 ^\tau dt ~ \text{tr}\left[(\partial_\alpha ^1 \partial_\beta ^2 \mathcal{L})_{\theta^1 = \theta^2 = \theta} \rho_{\text{ss}}  \right]
    \;.
\end{align}
Noting that $(\partial_\alpha ^1 \partial_\beta ^2 \mathcal{L})_{\theta^1 = \theta^2 = \theta} = \delta_{\alpha \beta} L_\alpha^{\theta_\alpha} \rho (L_\alpha^{\theta_\alpha})^\dagger /4$, the quantum Fisher information matrix becomes diagonal, with elements
\begin{align}
    [\mathcal{I}_Q(\tau)]_{\theta_\alpha \theta_\beta} = \tau \delta_{\alpha \beta} a_{\alpha}
    \;,
\end{align}
where $a_\alpha = {\rm tr}{\bm(} L_\alpha^{\theta_\alpha} \rho (L_\alpha^{\theta_\alpha})^\dagger{\bm)}$.
Substituting the quantum Fisher information matrix into the Cram\'er-Rao bound, $\sum_{\alpha\beta} R_{\theta_\alpha}(\tau)[\mathcal{I}_Q^{-1}(\tau)]_{\theta_\alpha\theta_\beta} R_{\theta_\beta}$, yields the quantum FRI \eqref{eq:FRI_Q}.

\subsection{Derivation of the identity $R_{B_{ij}}(\tau)/R_{F_{ij}}(\tau) = 2J_{ij}/a_{ij}$}
The key step in deriving Eq.~\eqref{eq:FRI_B2} is proving the identity $R_{B_{ij}}(\tau)/R_{F_{ij}}(\tau) = 2J_{ij}/a_{ij}$ for both current-like and state-dependent observables.
This identity can be derived using standard linear response theory~\cite{kubo2012statistical,gardiner1985handbook}.
We begin by considering a perturbation in the symmetric parameter $B_{mn}$ shared by the transition rates $W_{mn}$ and $W_{nm}$.
Suppose we perturb $B_{mn}$ to $B_{mn}+\Delta B$ at time $t=0$.
The transition rates and the probability distribution are then altered as $W_{ij}'=W_{ij} \bm{(}1+(\delta_{im}\delta_{jn} + \delta_{in}\delta_{jm})\Delta B \bm{)}$ and $p_i(t) = \pi_i + q_i (t) \Delta B$, respectively, up to linear order in $\Delta B$.
Substituting these relations into the master equation \eqref{eq:Markov_ori_dyna} yields the equation for $q_i (t)$ as follows:
\begin{align}
    \dot{q}_i (t) = \sum_{j (\neq i)} (W_{ij}q_j (t) - W_{ji}q_i(t)) + \delta_{im} J_{in} + \delta_{in} J_{im}
    \;.
\end{align}
By introducing a stochastic matrix $\mathsf{W}$, where diagonal components are given by $[\mathsf{W}]_{ii} = -\sum_{j(\neq i)} W_{ji}$ and off-diagonal components by $[\mathsf{W}]_{ij} = W_{ij}$, the linear differential equation for $q_i(t)$ can be solved as
\begin{align}\label{eq:q_expr}
    q_i (t) = \int_0 ^t dt' \sum_j [e^{\textsf{W}(t-t')}]_{ij}(\delta_{jm} J_{jn} + \delta_{jn} J_{jm} ) \;.
\end{align}
The time integration in Eq.~\eqref{eq:q_expr} can be performed using the spectral decomposition of the stochastic matrix $\mathsf{W}$, expressed as $[\mathsf{W}]_{ij} = \sum_\alpha \lambda_\alpha r_i^\alpha l_j^\alpha$, where $\lambda_\alpha$ are the eigenvalues of $\mathsf{W}$ and $l_i^\alpha~ ( r_i^\alpha)$ are the corresponding left  (right) eigenvectors normalized as $\sum_i l_i^\alpha r_i^\beta = \delta_{\alpha\beta}$.
The unique largest eigenvalue is $\lambda_0 = 0$, with corresponding eigenvectors given by $l_i^0 = 1$ and $r_i^0 = \pi_i$.
After carrying out the time integration and rearranging the terms, we obtain
\begin{align}
    q_i (t)  &= \sum_{\alpha (\neq 0)} \left(\frac{e^{\lambda_\alpha t} - 1}{\lambda_\alpha} \right) r_i ^\alpha (l_m ^\alpha -l_n ^\alpha)J_{mn}
    \;,
\end{align}
where the eigenvalues are indexed in descending order, i.e., $0 = \lambda_0 > \lambda_1 \geq \lambda_2 \geq \cdots$.
The dynamic response of state-dependent observables is then given by
\begin{equation}\label{eq:R_B_state}
\begin{aligned}
    R_{B_{mn}}(\tau) & = \int_0 ^\tau dt \sum_{i} g_i q_i(t) \\
    & = J_{mn} \sum_i g_i {\bm (}H_{im}(\tau) - H_{in}(\tau) {\bm )}
    \;,
\end{aligned}
\end{equation}
where the function $H_{ij}(\tau)$ is defined as
\begin{align}
    H_{ij}(\tau) \equiv \sum_{\alpha (\neq 0)} \left(\frac{e^{\lambda_\alpha \tau}- 1 - \lambda_\alpha \tau}{\lambda_\alpha ^2} \right)r_i ^\alpha l_j ^\alpha
    \;.
\end{align}
Since $\lim_{\tau\to 0} H_{ij}(\tau) = 0$, the kinetic response of state-dependent observables vanishes as $\tau\to 0$, as shown in Fig.~\ref{fig:fig_FRI_B2}(b).

When $B_{mn}$ is perturbed to $B_{mn} + \Delta B$ at time $t=0$, the current flowing between states $i$ and $j$ changes as $J_{ij} (t) = J_{ij} + K_{ij}(t) \Delta B$ up to linear order in $\Delta B$, where $K_{ij}(t) = (\delta_{im}\delta_{jn} + \delta_{jm}\delta_{in})J_{ij} + W_{ij} q_j(t) - W_{ji} q_i(t)$.
In the expression for $K_{ij}(t)$, the first term arises from the change in $W_{ij}$, while the second term results from the change in $p_i(t)$.
Substituting Eq.~\eqref{eq:q_expr} into $K_{ij}(t)$, we obtain
\begin{equation}
\begin{aligned}
    & K_{ij}(t) 
    =(\delta_{im}\delta_{jn}-\delta_{jm}\delta_{in})J_{mn} \\
    & + \sum_{\alpha(\neq 0)} \left(\frac{e^{\lambda_\alpha t} -1}{\lambda_\alpha} \right)(W_{ij}r_j ^\alpha -W_{ji} r_i ^\alpha)(l_m ^\alpha - l_n ^\alpha) J_{mn}
    \;.
\end{aligned}
\end{equation}
The dynamic response of current-like observables to the kinetic perturbation is then given by
\begin{equation}\label{eq:R_B_current}
\begin{aligned}
    & R_{B_{mn}}(\tau) = \int_0 ^\tau dt \sum_{i<j} \Lambda_{ij} K_{ij}(t) \\
    & = J_{mn}\left(\tau \Lambda_{mn} + \sum_{i\neq j} \Lambda_{ij} W_{ij} (H_{jm}(\tau) - H_{jn}(\tau)) \right)
    \;,
\end{aligned}
\end{equation}
Since $\lim_{\tau\to 0} H_{ij}(\tau)/\tau = 0$, the kinetic response of current-like observables $\mathcal{J}(\tau)$ satisfies $\lim_{\tau \rightarrow 0} R_{B_{mn}}(\tau)/\tau = J_{mn} \Lambda_{mn}$.
By noting that $\lim_{\tau\rightarrow 0} \text{Var}(\mathcal{J}(\tau))/\tau = \sum_{i<j} \Lambda_{ij} ^2 a_{ij}$, we confirm that the FRI \eqref{eq:FRI_B2} becomes an equality for current-like observables in the limit $\tau\rightarrow 0$ as shown in Fig~\ref{fig:fig_FRI_B2}(a). 

The analysis for a perturbation in the antisymmetric parameter $F_{mn}$ is similar to that for $B_{mn}$.
When $F_{mn}$ is perturbed to $F_{mn} + \Delta F$ at time $t=0$, the transition rates are altered as $W_{ij}' = W_{ij} [ 1 + (\delta_{im}\delta_{jn} - \delta_{in}\delta_{jm}) \Delta F/2 ]$ up to linear order in $\Delta F$.
The change of the sign in front of $\delta_{in}\delta_{jm}$ and the factor $1/2$ replace only $J_{mn}$ with $a_{mn}/2$ in Eq.~\eqref{eq:q_expr}.
This leads to replacing $J_{mn}$ with $a_{mn}/2$ in Eqs.~\eqref{eq:R_B_state} and \eqref{eq:R_B_current}, thereby proving the identity $R_{B_{mn}}(\tau)/R_{F_{mn}}(\tau) = 2J_{mn}/a_{mn}$ for current-like observables and state-dependent observables, and thus their linear combinations.
Using the same reasoning as for kinetic perturbations, we find that $\lim_{\tau\to 0}R_{F_{mn}}(\tau) = 0$ for state-dependent observables. Consequently, the FRI \eqref{eq:FRI_F} becomes an equality for current-like observables as $\tau \to 0$.

\end{document}